\documentclass[sigconf]{acmart}

\usepackage{xpatch}

\makeatletter
\xpatchcmd{\ps@firstpagestyle}{}{}{\typeout{Third patch succeeded}}{\typeout{Third patch failed}}

\xpatchcmd{\ps@standardpagestyle}{Manuscript submitted to ACM}{}{\typeout{Fourth patch succeeded}}{\typeout{Fourth patch failed}}   

\makeatother

\AtBeginDocument{%
  \providecommand\BibTeX{{%
    \normalfont B\kern-0.5em{\scshape i\kern-0.25em b}\kern-0.8em\TeX}}}

\copyrightyear{2024}
\acmYear{2024}
\setcopyright{rightsretained}
\acmConference[CHI EA '24]{Extended Abstracts of the CHI Conference on Human Factors in Computing Systems}{May 11--16, 2024}{Honolulu, HI, USA}
\acmBooktitle{Extended Abstracts of the CHI Conference on Human Factors in Computing Systems (CHI EA '24), May 11--16, 2024, Honolulu, HI, USA}
\acmDOI{10.1145/3613905.3651032}
\acmISBN{979-8-4007-0331-7/24/05}

\begin{document}

\title[Breaking the Plane]{Breaking the Plane: Exploring Real-Time Visualization of 3D Surfaces in Augmented Reality with Handwritten Input}


\author{Liam Esparraguera}
\authornote{All authors contributed equally to this research.}
\affiliation{%
  \institution{Princeton University}
  \city{Princeton, New Jersey}
  \country{USA}
}
\email{liame@princeton.edu}

\author{Brian Lou}
\authornotemark[1]
\affiliation{%
  \institution{Princeton University}
  \city{Princeton, New Jersey}
  \country{USA}
}
\email{blou@princeton.edu}

\author{Kris Selberg}
\authornotemark[1]
\affiliation{%
  \institution{Princeton University}
  \city{Princeton, New Jersey}
  \country{USA}
}
\email{kselberg@princeton.edu}

\author{Jenny Sun}
\authornotemark[1]
\affiliation{%
  \institution{Princeton University}
  \city{Princeton, New Jersey}
  \country{USA}
}
\email{jwsun@princeton.edu}

\author{Beza Desta}
\affiliation{%
  \institution{Princeton University}
  \city{Princeton, New Jersey}
  \country{USA}
}
\email{bd8405@princeton.edu}

\author{Andrés Monroy-Hernández}
\affiliation{%
  \institution{Princeton University}
  \city{Princeton, New Jersey}
  \country{USA}
}
\email{andresmh@princeton.edu}

\author{Parastoo Abtahi}
\affiliation{%
  \institution{Princeton University}
  \city{Princeton, New Jersey}
  \country{USA}
}
\email{abtahi@princeton.edu}
 

\begin{abstract}
We introduce Breaking the Plane, an augmented reality (AR) application built for AR headsets that enables users to visualize 3D mathematical functions using handwritten input. Researchers have demonstrated that overlaying 3D visualizations of mathematical concepts through AR increases motivation to learn mathematics and enhances mathematical learning, while also revealing that equation parsing via optical character recognition (OCR) makes the authoring of teaching materials more accessible and time-efficient for instructors. Previous works have developed AR systems that separately employ equation parsing and 3D mathematical visualizations, but work has yet to be done to combine those features by enabling real-time interactions and dynamic visualizations that help users learn in situ. We address this issue by developing an interactive AR system featuring handwritten equation parsing, object manipulation, and a custom 3D function plotter. We evaluated our system using a within-subjects study wherein 10 participants compared our system to two other commonly used 3D visualization systems: GeoGebra Desktop and GeoGebra AR. We found Breaking the Plane significantly surpassed other tools in engagement, achieved comparable ease of use to GeoGebra Desktop, and was rated as the most effective in aiding problem-solving, with a strong preference among participants for future use.
\end{abstract}



\begin{teaserfigure}
   \includegraphics[width=1\textwidth]{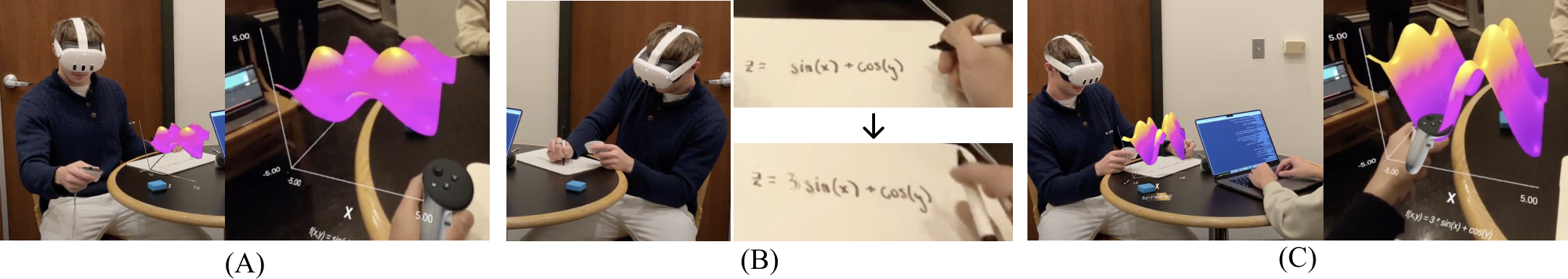}
   \caption{Images during use of the Breaking the Plane system. (A) The system scans the existing equation on the whiteboard \( z = \sin(x) + \cos(y) \) and renders a 3D visualization of the graph in augmented reality (B) The user modifies the equation on the whiteboard to \( z = 3\sin(x) + \cos(y) \) (C) The headset scans the new equation in real-time and renders a 3D visualization of the graph that the user can manipulate by grabbing, rotating, scaling, moving axes, etc.}
   \Description
   \label{fig:teaser}
\end{teaserfigure}

\maketitle
\section{Introduction}
For many students, the development of an intuitive understanding of multidimensional mathematical objects, such as multivariate functions and quadric surfaces, has proven a challenge \cite{gutierrez2004characterization}. This is in part due to the inherent difficulty in mentally visualizing abstract mathematical equations (e.g., $z=x^2+y^2$) as geometric objects and representing these objects on a 2D surface (e.g., paper, blackboard) \cite{ismail2020investigating}. 

Researchers have proposed AR technology as a solution to help students better understand 3D mathematical objects, and AR systems have demonstrated potential to increase motivation and improve learning outcomes in mathematics education \cite{bulut2023systematic}. Recent work has developed mobile AR interfaces that utilize an equation parsing component to enable existing teaching materials to be adapted for AR use through real-time visualization of 2D functions, creating a more interactive learning experience \cite{chulpongsatorn2023augmented}. We extend this work to a head-mounted display (HMD) system featuring real-time handwritten equation parsing. This approach leverages the unique strengths of AR in aiding students' comprehension and engagement in 3D mathematics and problem-solving, while also offering the dynamic responsiveness and hands-free control of visualizations made possible by this novel input method. Thus, the user can benefit from computer-aided math visualizations without needing to context-switch between apps or devices while problem-solving.

In this paper, we design and discuss Breaking the Plane, an AR headset application that uses real-time equation recognition, enabled by a Wizard-of-Oz implementation, and a 3D function plotter to improve ease of use and engagement when assisting students in the visualization of 3D mathematical objects. We evaluate our system on effectiveness of engagement, ease of use, and enhancement of problem-solving abilities relative to several other visualization techniques commonly used in math pedagogy. We found that participants using our system felt much more engaged with conceptual mathematical problems, experienced better ease of use than with an existing mobile AR system (GeoGebra AR), and were likely to use the system again in the future.
\section{Related Work}
\subsection{Augmented Reality in Mathematics Education} 
Researchers have demonstrated that AR systems in mathematics education which assist students by visualizing functions, vectors, and equations in 3D space enhance mathematical learning, contribute to memory retention, and increase motivation to learn mathematics, among other benefits \cite{chang2022ten, gargrish2021measuring, liono2021systematic, salinas2013development}. Consequently, AR technology has been proposed as a solution to help students better understand 3D mathematical objects but has yet to be widely adopted in classrooms, despite having been shown to improve learning outcomes in mathematical education \cite{da2019don}. A survey of 106 teachers found that the second most frequently stated blocker for adopting AR in the classroom was "I prefer that my
students create AR rather than use pre-created programs"  \cite{da2019don}. A systematic review also found that one of the most frequent disadvantages of AR in mathematics education is the difficulty of developing materials through AR \cite{palanci2021does}. An AR application with real-time equation parsing would enable students to author personalized visualizations to support their learning. 
\subsection{Current Approaches to Mathematics Visualization}
\subsubsection{Plotting Graphs in 3D Space}
Existing AR systems have been developed to address the challenges in visualizing equations. GeoGebra provides a system to plot equations without AR on desktops (GeoGebra Desktop) and with AR on mobile (GeoGebra AR) \cite{celen2020student}. GeoGebra AR is a mobile app that allows users to input equations manually and visualize their corresponding 3D graphs in a mixed-reality space with AR. Researchers evaluated the effectiveness of this app in aiding students and found the system to be practical and effective, resulting in students having reported improved visual-spatial understanding after using the tool \cite{mailizar2021exploring}. However, in this study, the small size of smartphone screens was cited as a limitation, as users could not see much detail in the augmented graphs, suggesting a lack of ease of use \cite{mailizar2021exploring}. 

Alternatively, GeoGebra Desktop is a widely adopted desktop app that allows users to manually input equations and visualize 3D equations on a traditional computer screen \cite{celen2020student}. However, in a previous study, students reported low confidence in using GeoGebra Desktop, which was hypothesized to be due to insufficient time dedicated to learning keystrokes \cite{shadaan2013effectiveness}. Equation parsing would eliminate the need for learning keystrokes. 

\subsubsection{Equation Parsing in AR}
To address this problem, researchers recently developed an AR interface for mathematical equations adding OCR input \cite{chulpongsatorn2023augmented}. The interface allows users to point a mobile device camera at an equation in a textbook, which the system parses and generates an interactive plot which is then overlaid on the device's display. This study found that such a system created a more engaging learning experience for students compared to traditional approaches and reduced the necessary authoring effort for AR learning materials, allowing for existing textbook resources to be programmatically extended into interactive AR experiences. Therefore, students do not have to learn specific keystrokes to input equations and only have to point their phone camera at a page. However, this system does not support equation parsing of handwritten input or visualization of 3D surfaces.
 
\subsection{A Novel Approach}
Past approaches have separately used equation parsing and visualization of 3D equations in the physical space with AR to address the issue of aiding mathematical understanding; however, work has yet to be done to combine the two. Such a synthesis holds value, as an AR system that supports the generation of 3D mathematical visualizations based on print or handwritten material can both take advantage of the unique strengths of AR systems in improving individuals’ ability to interpret 3D geometry and the ability of equation parsing to reduce the friction of information input and content authoring. Thus, we propose a novel system that extends 3D AR visualization tools with handwritten equation-parsing input as a tool to understand multidimensional mathematical objects.

\begin{figure}[h!]
\vspace{-2pt}
\includegraphics[width=0.45\textwidth]{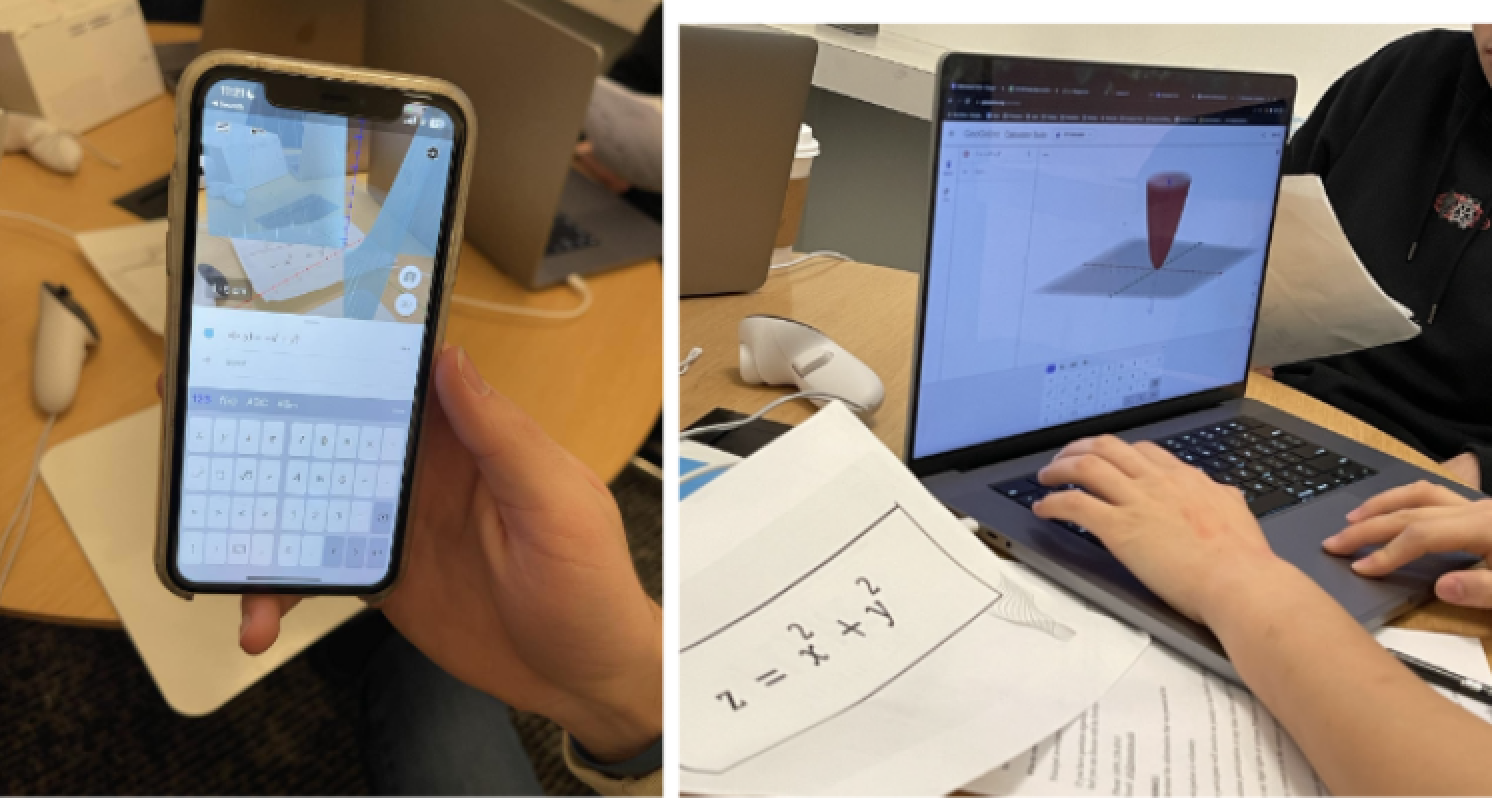}
  \includegraphics[width=0.45\textwidth]{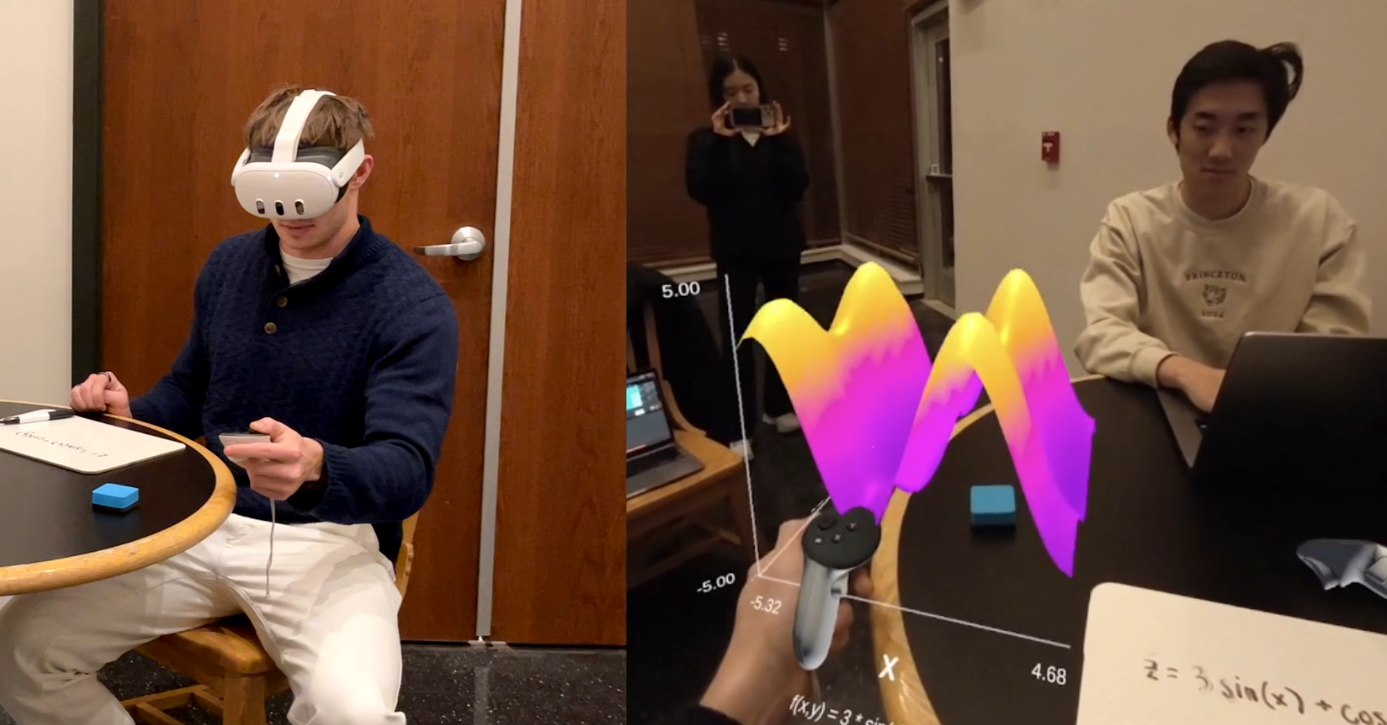}
  \caption{Our technical systems in the order: GeoGebra AR, GeoGebra Desktop, Breaking the Plane (our system on Meta Quest 3)}
  \label{fig:diffsystems}
\end{figure}
\vspace{-2pt}
\section{Implementation}
Our vision is to enable an intuitive, hands-free experience in which users can simply wear a headset while studying multivariable equations and work alongside a hassle-free aid for mathematical visualization that extends their mental model. When a user encounters an equation, either in printed text or written text, the headset automatically detects the equation and renders its 3D graph in the user’s field of view. Users can then interact with these graphs: rotating, inspecting from various angles, and editing equations to see real-time changes in 3D space.

We developed our final system for the newly released Meta Quest 3 using Unity3D. The Quest 3 is an AR-capable headset featuring color passthrough and two controllers to allow for interaction with virtual objects. Via the Quest 3, users can see their physical surroundings while viewing and interacting with virtual graphs in full stereoscopic depth. 
\subsection{3D Function Plotter}
To visualize mathematical equations as interactive objects in an augmented reality (AR) scene, our system requires a component that can parse a user-input mathematical function and output a three-dimensional surface representation of the function as a mesh model. Due to the lack of existing open-source resources providing such functionality, we elected to implement this component of our system as a custom script that accepts as input a string of characters to be parsed as a function of two variables (x, y) and generates a procedural mesh that plots the output of the parsed function over a user-specified domain with a variable sampling density. To interpret the text representation of the user’s function, we utilize NCalc, a library for mathematical expression evaluation, to calculate the numerical values of the string-formatted function over the input domain. 

To constrain the visible bounds of the plotted function and better convey the relative value of the function across its input domain, we wrote a shader that uses the rendered pixel’s height in local object coordinates to determine its color using a custom colormap gradient and restrict the surface to a variable height range. Finally, the graph is displayed with labeled axes and a text representation of the plotted function to inform the user of the current input domain and mathematical expression of the visualization.
\raggedbottom
\subsection{Manipulation of AR Graph}
We implemented controls for the graph’s input and output domains that allow the user to better manipulate, and hopefully understand, the visualized 3D mathematical surface. To allow the user to move the AR graph about their physical surroundings, our system makes use of the Quest 3’s Touch controllers to permit the user to grab the graph and move it around using the secondary hand trigger. In addition, to translate and scale the plotted function within the graph, the user can pan and zoom in/out the input domain (the x and y axes) using the thumbstick and A/B buttons, respectively; these same controls modify the z-axis limits when the user holds the primary trigger. Finally, to allow the user to easily re-center and re-scale the graph to its initial state, one can click the thumbstick to reset all axis parameters to their default values. 
\subsection{Parsing Handwritten Equations}
The API of the Meta Quest 3 headset prevents external app developers from accessing, viewing, or storing direct images from the device's cameras. Thus, building our system on the Meta Quest 3 platform meant that we could not use a typical OCR methods while using this hardware, as any automated text-recognition technique would require access to the headset’s camera feed.

Given this limitation, we decided to adopt a “Wizard-of-Oz” approach to equation parsing in which the user wears the headset and an operator (a member of our team) views their streamed live camera feed from the Quest device and manually inputs the equation to be rendered on the graph. We established a WebSocket connection between our local laptop and the headset to facilitate real-time communication between the two. To this end, we also implemented a command that allows the “Wizard” to broadcast a loading indicator stating “OCR Processing…” to the user, thus disguising operator delay by the appearance of a plausible computation time for equation parsing and granting the end user a more consistent indicator of the system’s state. Overall, this method supports  a user experience comparable to a system with working OCR, has the added benefit of functioning without explicit input by the user for scanning, and is likely to offer a better degree of equation transcription accuracy in user studies.

Previous solutions have focused on allowing users to manually input equations for improved visualization and manipulation. We believe that our simulation of dynamic equation parsing improves a core problem of ease of use with AR-based approaches to mathematical learning by allowing for automatic, hands-free input and more effective manipulation. 
\section{Evaluation}
\subsection{Study Participant Demographics}
To evaluate our system, we recruited 10 adult participants (self-identified as 6 male, 4 female, aged 18-21), all of whom are students from our institution who have formerly taken or are currently taking an introductory multivariate calculus course. We recruited most of the participants by sending out emails to the student body providing brief information about the study and advertising a \$15 payment as an incentive in compensation for their participation. This was our recruiting method with the exception of our last study participant, who we recruited by asking students at our institution's engineering library whether they had taken multivariate calculus and were interested in participating in a study; we chose this approach since the original study participant mistakenly signed up despite having not taken a multivariate calculus course. 

\subsection{Study Procedure and Data Collection Methodology}
For our study, we employed a within-subject evaluation design in which each participant tests four systems for mathematical visualization. Each session lasted $\sim$45 minutes, and our team had a planned-out routine and script to follow to keep each study as similar as possible. After giving a brief introduction and biographical questions, participants underwent four rounds. In each round, we assigned a quadric surface and a system to the user, both of which we selected in independent random sequences to counteract order-dependent effects.
The system choices were: (1) No system/technology, whiteboard and marker only, (2) GeoGebra AR (iOS application), (3) GeoGebra Desktop (Web Application), (4) Breaking the Plane (our system).

\subsubsection{Evaluation Study Questions}
For each of the four random pairings of the above visualization systems and 3D surface types, we asked each participant a series of questions that would test their ability to make predictions about the relationship between the mathematical function and its geometric and spatial properties. Note that these questions were not graded for ‘correctness’ – their sole purpose was to prompt the user to (optionally) engage with the provided visualization system, should they find it potentially useful. See Appendix for the questions.

\subsubsection{Post-Demonstration Questionnaire}
After all four rounds of system demonstration, we asked each study participant to fill out a post-interview questionnaire about their specific experiences with each system and overall feedback on our study. Questions regarding participants’ experiences with the four visualization systems types were formatted as 5-point Likert scale responses rating each system’s perceived ease-of-use, user engagement, and the participant’s willingness to use the system again, to facilitate statistical comparisons of these properties across systems. In addition, the questionnaire required the user to rank the four systems in order of their relative effectiveness in solving the provided problems, and also included 5-point Likert scale questions directly comparing components of our headset AR system, such as input via handwritten equations and 3D AR visualizations, to those of the mobile AR and desktop systems. Finally, several optional, open-ended questions prompted the user to share their thoughts on the four systems’ ease of use, as well as the design and execution of the study.

\section{Study Results}
The study's results indicate that while our AR headset system and GeoGebra Desktop are perceived similarly in terms of ease of use, our system significantly outperforms GeoGebra AR, No System, and GeoGebra Desktop in engagement, with participants showing a marked preference for our system due to its interactive AR features and equation parsing functionality. Furthermore, Breaking the Plane was most often ranked as the most effective tool for solving mathematical problems, with participants expressing a strong likelihood of adopting our system in the future due to its user-friendly interface and superior functionality in educational contexts.

\subsection{Ease of Use}
The t-test results provided in Fig. \ref{fig:easeofuse} demonstrate significant variances in ease of use among different systems. One participant pointed to this in their response, "The mobile app was the hardest learning curve to grasp since it was such a small screen and not really able to move the graph," reflecting the lower perceived ease of use for the GeoGebra AR and no system. While comparisons between GeoGebra Desktop and GeoGebra AR, as well as between GeoGebra AR and our system, showed statistically significant differences (p-value <  0.05), there was no significant disparity in ease of use between No System and GeoGebra AR, or GeoGebra Desktop and our system, indicating a comparable ease of use for our system and GeoGebra Desktop and a clear user preference for these two over GeoGebra AR and the lack of any computer-generated visualization.

\subsection{Engagement}
\begin{figure}[h!]
\vspace{-2pt}
\includegraphics[scale=0.25]{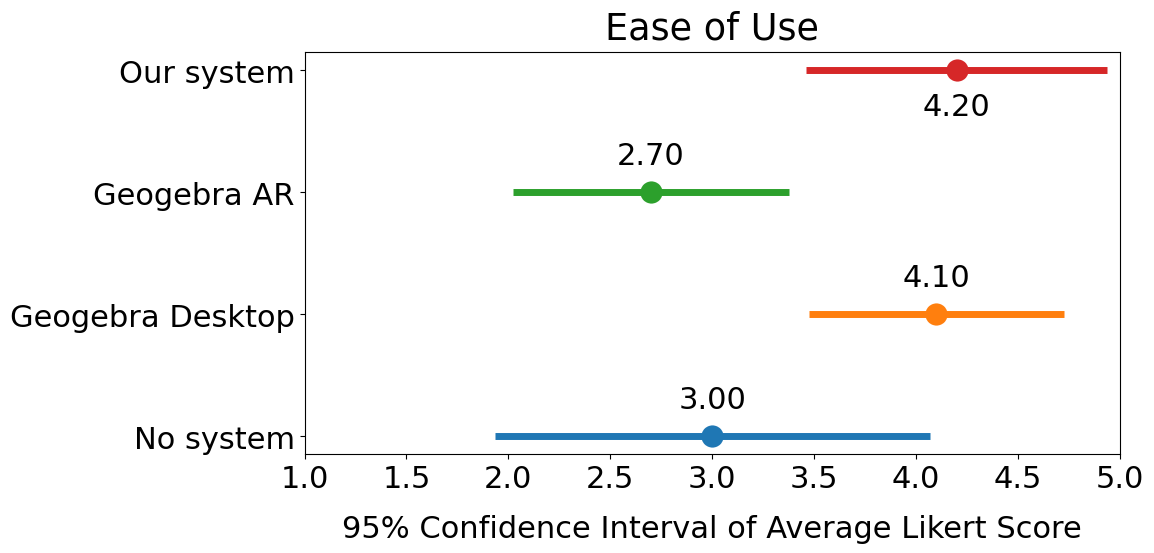}
  \includegraphics[scale=0.25]{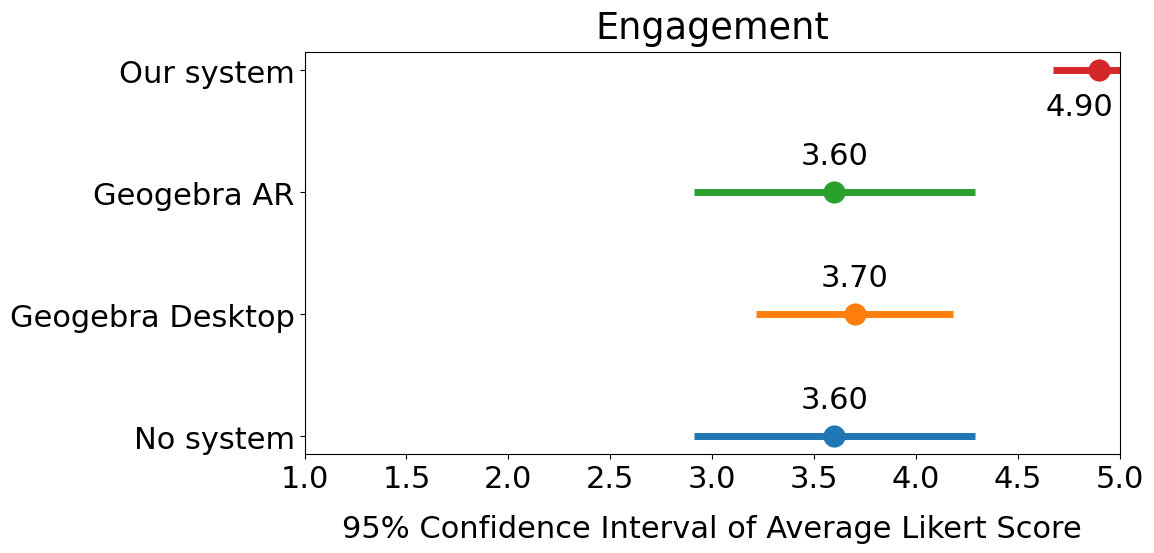}
  \caption{Graphs comparing user response on the perceived levels of engagement and ease of use with different systems. Likert-scale responses are converted to an integer (1-5) scale and displayed with a 95\% confidence interval.}
  \label{fig:easeofuse}
\end{figure}

The paired t-tests in Fig. \ref{fig:easeofuse} indicated clear differences in engagement levels across conditions. Participants demonstrated statistically significantly higher engagement with "Our System" compared to all the other systems: "No System" (t-statistic = -3.667, p = 0.008 < 0.05), "GeoGebra Desktop" (t-statistic = -5.000, p = 0.004), and “GeoGebra AR” (t-statistic = -3.057, p = 0.022) suggesting a notable increase in engagement when using our headset-based AR system. The 95\% confidence intervals for mean engagement scores displayed below not only confirmed the precision of these differences but also underscored the effectiveness of our system in augmenting educational problem-solving engagement.

\vspace{-2pt}
\subsection{Effectiveness in Solving Problems}
Of the four systems tested, our AR headset system was most frequently ranked as the most effective (6 times) and second most effective (4 times) in aiding mathematical problem-solving during the study. The GeoGebra Desktop system was also perceived positively, with several rankings in the top two positions. The other systems were more commonly ranked as less effective, with the Mobile AR System most often ranked third and no system most often ranked as the least effective.
\begin{figure}[h!]
    \includegraphics[scale=0.175]{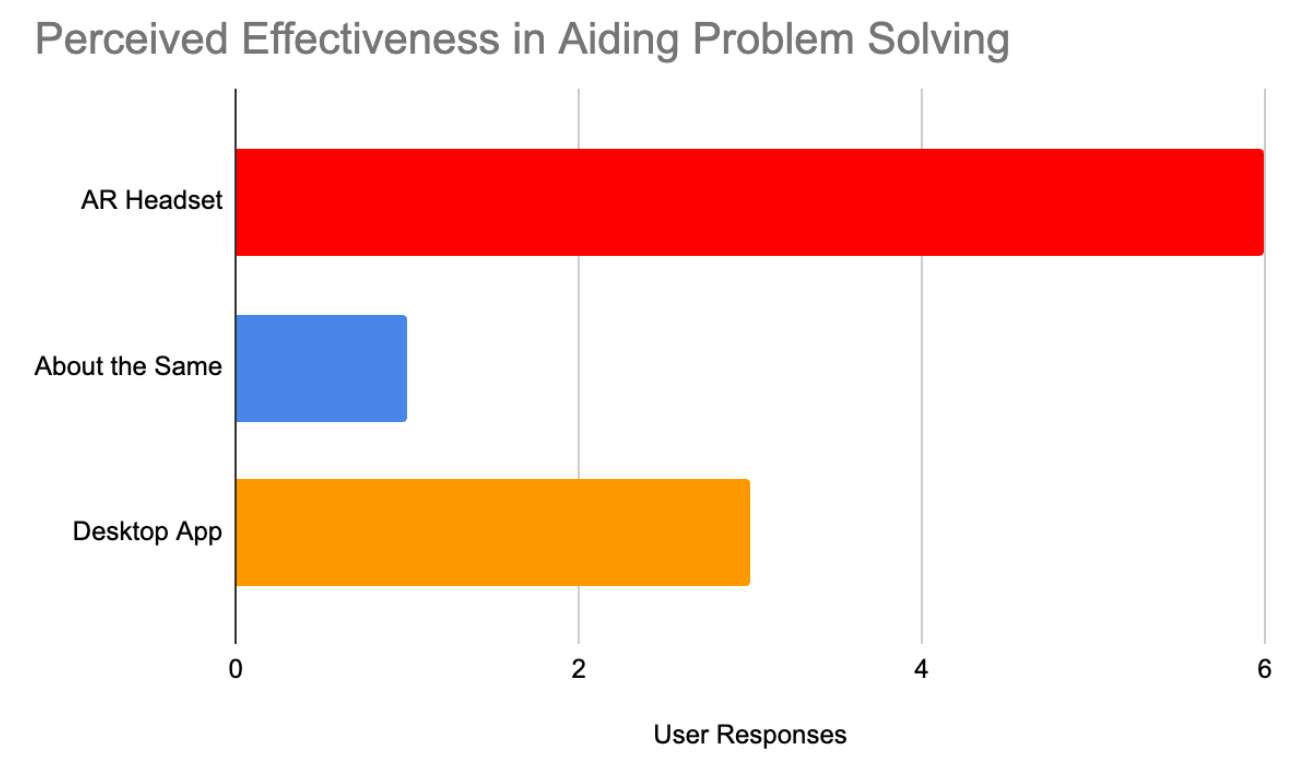}
  \includegraphics[scale=0.175]{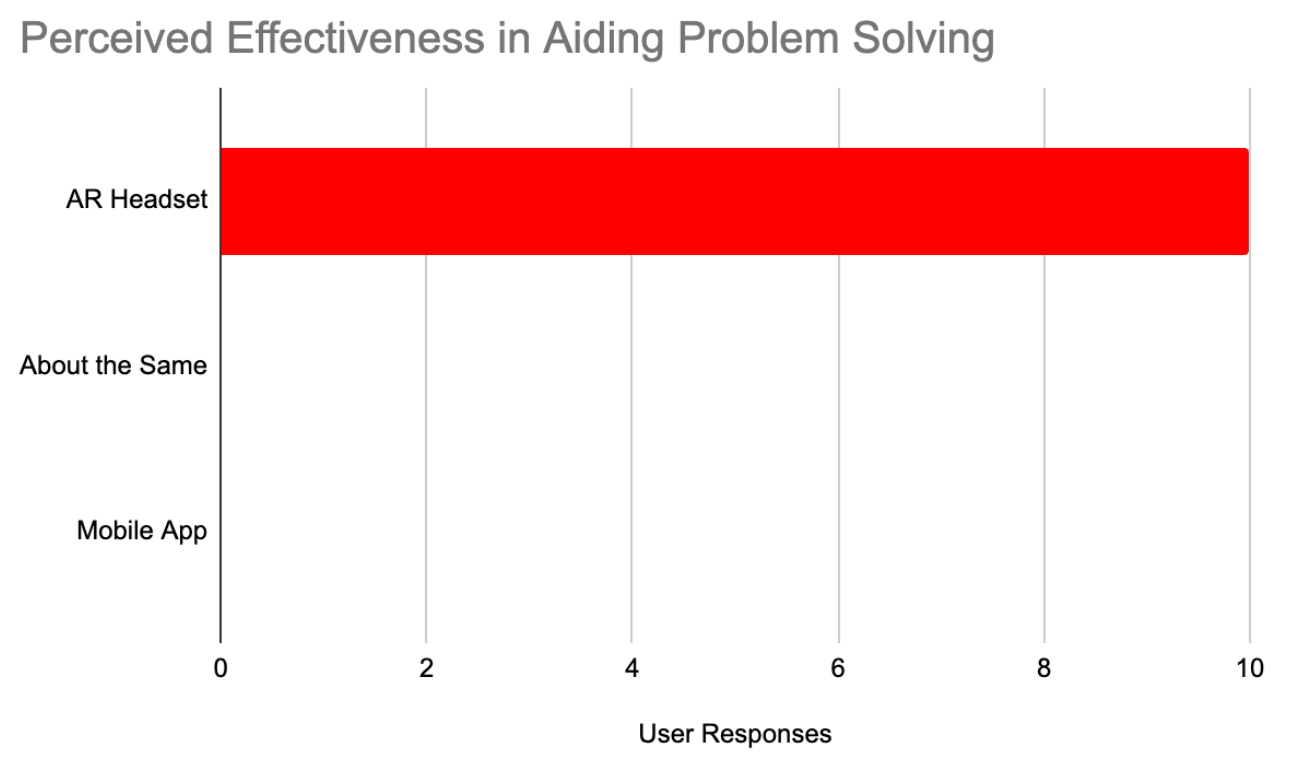}
  \caption{Direct comparison of our system with GeoGebra Desktop (Desktop App), GeoGebra AR (Mobile App) in aiding problem-solving}
  \label{fig:comparisonwithothersystems}
\end{figure}
\subsection{Satisfaction/Future Adoption of Our System}
In the final survey, participants expressed high satisfaction with our AR system, particularly valuing its intuitive graph manipulation and handwritten equation parsing capabilities. For example, P7 stated that "The headset was the most user-friendly as I was able to physically use my hand to grab and manipulate the graph/surface, while on a desktop or mobile app, those motions didn't always directly translate to what I wanted to do with the graph". This satisfaction was quantified with a high mean score of 4.8 and a standard deviation of 0.42 (on a 5-point Likert scale) for the AR headset's features, significantly surpassing the mobile app's mean score of 2.3 and a standard deviation of 0.67. Additionally, the likelihood of future use was strong, reflected in a mean score of 4.3 for potential reuse of the AR headset. The contribution of input via handwritten equations to problem-solving was also highly rated, with a mean score of 4.6. These statistics underscore the system's superior functionality and user experience, suggesting its potential for future adoption in educational settings.

\section{Discussion}
Our data analysis demonstrates a significant user preference for our system over existing methods for mathematical visualization for the tasks presented, most notably in comparison to GeoGebra AR, a prominent existing AR system (Fig. \ref{fig:comparisonwithothersystems}). The findings of our study strongly support the hypothesis that the novel features introduced by our system --- headset-based AR and handwriting input via equation parsing --- enhance the user experience by significantly improving (relative to GeoGebra AR) ease of use, user engagement, and willingness to reuse the AR headset. For example, P2 expressed, “I found it somewhat difficult at first to use the AR headset since I had never used it before, but was able to adapt to it.” P9 noted “[the handwritten input allowed me] to write down the equations instead of having to learn the syntax for typing them into the desktop or mobile app”. This ease of use stems from the fact that conventional systems demand users to familiarize themselves with system-specific syntaxes for entering equations. However, our system utilizes handwritten input, which accommodates any syntax a user intuitively uses and eliminates the need for this additional learning --- interactivity is at the rate of the user's handwriting. 

The enhanced engagement provided by our AR system is critical in educational contexts, as the ability to attract and maintain user interest is often correlated with better learning outcomes. In addition, our system and GeoGebra Desktop were found to be comparably user-friendly, suggesting that AR technology can be as intuitive as traditional desktop applications in mathematical learning contexts given effective interface design. Nevertheless, some users still preferred the Desktop app, potentially due to prior use and the familiarity of mouse-and-keyboard interfaces. We think there may be a tradeoff between some users’ preferences for a system that is already familiar to them and easy to access versus one that is holistically more intuitive but has a slight learning curve. Further research with larger sample sizes and diverse educational contexts would be beneficial to fully understand the implications of AR in education.

\section{Limitations}
Our findings are limited by several factors including the use of a Wizard-of-Oz (WoZ) approach to equation parsing, a limited sample size of participants from a homogeneous population, and the influence of potential learning effects on our study. The choice of a WoZ solution using manual input for our system’s equation parsing component reduced the applicability of our findings to future systems using fully-functional OCR methods and introduced variability to our evaluation due to inconsistencies in operator performance: there was significant variance in system response times to new user input, which may influence perceived ease of use and utility. Resulting from our within-subjects design, we also observed the self-reporting of a learning effect within our study wherein several participants noted that they felt the conceptual questions in the final rounds of our study to be easier than those in the early rounds, as they had remembered results and visualizations from previous systems. To mitigate this effect, future studies should select conceptual questions and multivariable equations from a larger set of potential options, ensuring that participants do not receive repeated or similar material. Most significantly, the sample size of our study was limited (n = 10) and was drawn from a small, relatively homogeneous demographic: undergraduate students who had previously taken a multivariate calculus course at our institution. To address the utility of AR for mathematical pedagogy in a variety of educational contexts, future work should explore the responses to pedagogical AR systems of diverse student populations.

Our study involved undergraduate participants from our institution, none of whom had reported vision, hearing, or haptic disabilities. Additionally, the Quest 3 headset used in our study poses challenges for users with visual impairments or motor control issues. Recognizing this, we propose future research to explore alternative control methods in AR environments, such as voice commands or adaptive controllers. 
\vspace{-5pt} 
\section{Conclusion and Future Work}
Our research delves into the potential of 3D AR visualizations to enhance the understanding of multidimensional mathematical concepts. We successfully implement and evaluate an AR headset system equipped with handwritten equation-parsing input to facilitate a deeper and more intuitive comprehension of complex mathematical objects. Our findings indicate that our system, compared to mobile AR and traditional flat-screen visualizations, significantly improves user engagement and rivals the ease of use of a desktop-based application, GeoGebra Desktop.

Key insights from our study reveal that participants not only found our system to be highly engaging due to its interactive AR features but also appreciated the equation parsing functionality for its ease of input. This level of engagement is crucial in educational settings where user interest and exploration are closely linked to learning outcomes. Moreover, despite initial adjustment challenges, users quickly adapted to the system, highlighting its potential for wider adoption in educational environments. However, our research also encounters certain limitations, including the reliance on a Wizard-of-Oz approach for handwriting input, a study demographic bounded to multivariate calculus students, and a small sample size. These factors limit the generalizability of our findings and suggest the need for further research with diverse populations and practical OCR implementations.

Looking ahead, we envision exploring alternative input methods in AR environments, such as voice commands or gesture recognition, to make our system more accessible to users with different abilities, as well as the potential for multi-user collaboration. In conclusion, our study underlines the transformative potential of AR in educational settings, particularly in mathematical education. Future research should focus on exploring the benefits of HMD AR technology with handwritten input --- increasing engagement, enabling rapid and self-directed learning, and simplifying complex concept visualization --- in mathematics education and other domains to facilitate better learning outcomes and wider adoption of AR in modern education.


\bibliographystyle{ACM-Reference-Format}
\bibliography{refs}
\end{document}